%% file: semcom.tex
\documentclass[dvips,envcountsect
]{llncs}

\input{macros}

\title{Bicompletions of distance matrices\\[2ex]
{\large \it To Samson Abramsky on the occasion of his 60th birthday}}

\author{Dusko Pavlovic\\
\email{\small Email:~dusko.pavlovic@rhul.ac.uk}
}
\institute{Royal Holloway, University of London, and University of Twente}

\date{}

\begin{document}
\maketitle

\begin{abstract}
In the practice of information extraction, the input data are usually arranged into \emph{pattern matrices}, and analyzed by the methods of linear algebra and statistics, such as principal component analysis. In some applications, the tacit assumptions of these methods lead to wrong results. The usual reason is that the matrix composition of linear algebra presents information as flowing in waves, whereas it sometimes flows in particles, which seek the shortest paths. This wave-particle duality in computation and information processing has been originally observed by Abramsky. In this paper we pursue a particle view of information, formalized in \emph{distance spaces}, which generalize metric spaces, but are slightly less general than Lawvere's \emph{generalized metric spaces}. In this framework, the task of extracting the 'principal components' from a given matrix of data boils down to a \emph{bicompletion}, in the sense of enriched category theory. We describe the bicompletion construction for distance matrices. The practical goal that motivates this research is to develop a method to estimate the hardness of attack constructions in security.  
\end{abstract}

\section{Introduction}\label{Introduction}
\subsubsection*{Dedication.}
When Samson Abramsky offered me the position of 'Human Capital Mobility Research Fellow' in his group at Imperial College back in 1993, I was an ex-programmer with postdoctoral experience in category theory. It was a questionable investment. Category theoretical models of computation were, of course, already in use in theoretical computer science; but the emphasis was on the word 'theoretical'. A couple of years later, I left academia to build software using categorical models. While it is clear and well understood that  Samson's work and results consolidated and enriched categorical methods of theoretical computer science, their applications in the practice of computation may not be as well known. In the long run, I believe, the impact of the methods and of the approach that we learned from Samson will become increasingly clear, as the abstract structures that we use, including the fully abstract ones, are becoming more concrete, more practical, and more often indispensable.

In the present paper, I venture into an extended exercise in enriched category theory, directly motivated by concrete problems of security \cite{PavlovicD:FAST10,PavlovicD:NSPW11} and of data analysis \cite{PavlovicD:ICFCA12}. Although the story is not directly related to Samson's own work, I hope that it is appropriate for the occasion, since he is the originator of the general spirit of categorical variations on computational themes, even if I can never hope to approach his balance and style.

\subsection*{Motivation: Distances between algorithms} 
Suppose that you are given an algorithm  $a$, and you need to construct another algorithm $b$, such that some predicate $P(a,b)$ is satisfied. Or more concretely, suppose that $a$ is a software system, and $b$ should be an attack on $a$, contradicting $a$'s security claim by realizing a property $P(a,b)$. Since reverse engineering is easy \cite{BarakB:noobfus,GoldwasserS:noobfus}, we can assume that the code of $a$ is readily available, and your task is thus to code the attack $b$. Note that $a$ is in principle an algorithmic pattern, that can be implemented in many ways, and may have many versions and instances. So your attack $b$ should also be an algorithmic pattern, related to $a$ by some polymorphic transformation. The derivation of $b$ from $a$ should thus be polymorphic, i.e. a uniform construction: it should be a program $p$ that inputs a description of $a$ and outputs a corresponding description $p(a) = b$. How hard is it to find $p$? An approach to answering such questions is suggested in algorithmic information theory \cite{Levin-Zvonkin,Vitanyi:book}. The notion of \emph{Kolmogorov complexity}\/  is that the distance from an algorithm $a$ to an algorithm $b$ can be measured by the length of the shortest programs that construct $b$ from $a$, i.e. 
\bea\label{solomon}
\dis a b & = & \bigwedge_{p(a) = b} |p|
\eea
where $|p|$ denotes the length of the program $p$. It is easy to see that the above formula yields the triangle law $\dis a b + \dis b c \stackrel + \geq \dis a c$, where the superscript '+' means that the uniform order relation $\geq$ is taken up to a constant, which is in this case the length of the program composition operation, needed to get a program to construct $c$ from $a$ by composing a program that constructs $c$ from $b$ with a program that constructs $b$ from $a$. Algorithmic information theory always works with such order relations \cite{Vitanyi:book,Downey:book}. The equation $\dis a a \stackrel + = 0$ holds in the same sense, up to the constant length of the shortest identity program, that just inputs and outputs identical data. This distance of algorithms, in the style of Kolmogorov complexity, was proposed in \cite{PavlovicD:NSPW11} as a tool to measure how hard it is to construct an attack on a given system. The point was that a system could  be effectively secure even when some attacks on it exist, provided that these attacks are provably hard to construct. The goal of the present note is to spell out some general results about distance that turn out to be needed for this particular application.

But why do we need general results about distances to answer the concrete question about the hardness of constructing attack programs from system programs? The reason is that the task of finding an attack algorithm not too far from a system algorithm naturally leads to the task of construcing a \emph{completion}\/ of the space around the system algorithm. The attacker sees the system, and may be familiar with some other algorithms in its neighborhood; but it is not known whether an attack exists, and how far it is. The task of discovering the attack is the task of completing the space around the system.  And the construction of a completion is easier in general, than in some concrete cases. 

How does a real attacker search for an algorithm $p$ to derive an attack $b$ from the system $a$? He is not trying to guess the construction in isolation, but in the context of his algorithmic knowledge. This knowledge has at least two components. On one hand, there is some algorithmic knowledge $A$ about the software systems $a_0, a_1, a_2\ldots$, and a distance measure $A\times A\tto{d_A} [0,\infty]$ between them, which express how they are related with each other. On the other hand, there is some algorithmic knowledge $B$ about the attacks $b_0, b_1, b_2\ldots$, and their distances $B\times B\tto{d_B} [0,\infty]$. Last but not least, there is some knowledge which attacks are related to which systems. This knowledge is expressed as a distance matrix $A\times B \tto\Phi [0,\infty]$, where  shorter distances suggest easier attacks. 
In order to determine whether there are any attacks in the proximity of a given system $a$, our task is to conjoin  the distance space $A$ of systems with the distance space $B$ of attacks consistently with the distance matrix $A\times B \tto\Phi [0,\infty]$ where the observed connections between the systems and attacks are recorded. In this conjoined space, we need to find the unknown attacks close to the target system. We find them by completing the space of the known attacks. But since the completion is in general an infinite object, we first study it abstractly, to determine how to construct just the parts of interest.

\paragraph{Related work.} The completions that we study are based on Lawvere's view of metric spaces as enriched categories  \cite{LawvereFW:metric}. Lawvere's generalized metric spaces were extensively used in denotational semantics of programming languages \cite{WagnerK,RuttenJ:metric,Schellekens:Yoneda}, and recently in ecology \cite{LeinsterT:species}, following a renewed mathematical interest in the enriched category approach \cite{LeinsterT:magnitude}. In my own work, closely related results arose in the framework of information extraction and concept analysis \cite{PavlovicD:ICFCA12}. That work was, however, not based on distance spaces as categories enriched in the additive monoid $[0,\infty]$, but on \emph{proximity spaces}, or \emph{proxets}, as categories enriched in the multiplicative monoid $[0,1]$. Proxets are a more natural framework for concept analysis, because they generalize posets, as categories enriched over the multiplicative monoid $\{0,1\}$, and the existing theory and intuitions are largely based on posets. Distance spaces, on the other hand, appear to be a more convenient framework for relating algorithms.

%
%
%

\paragraph{Outline of the paper.} In Sec.~\ref{distance spaces-sec} we define distance spaces and describe some examples. In Sec.~\ref{sequences-sec} we spell out the notions of limit in distance spaces, the  basic completion constructions, and the adjunctions as they arise from the limit preserving morphisms. In Sec.~\ref{Matrices-sec}, we introduce distance matrices, and describe their decomposition. In Sec.~\ref{Semantic-compl-sec} we put the previously presented components together to construct the bicompletions of distance matrices.  Sec.~\ref{Conclusions-sec} provides a summary of the obtained results and a discussion of future work. 

\section{Distance spaces}\label{distance spaces-sec}
\subsection{Definition and background}\label{definition}
\begin{defn}\label{def-distance}
A\/ \emph{distance space} is a set $A$ with a \emph{metric} $d_A :A\times A\to [0,\infty]$ which is
\begin{itemize}
\item reflexive: $\dis x x = 0$,
\item transitive: $\dis x y + \dis y z\ \geq\ \dis x z$, and
\item antisymmetric: $\dis x y = 0 = \dis y x \ \Longrightarrow\  x=y$
\end{itemize}
A \emph{contraction}\/ between the distance spaces $A$ and $B$ is a function $f:A\to B$ such that for all $x,y\in A$ holds $\dist A x y   \geq  \dist B {fx} {fy}$.
The category of distance spaces and contractions is denoted $\Dist$.
\end{defn}

\paragraph{Background.} In topology, distance spaces have been studied since the 1930s under the name \emph{quasi-metric spaces} \cite{Wilson:quasi-metric,Kelly:bitopological}. The prefix 'quasi' refers to the fact that the metric symmetry law $d(x,y) = d(y,x)$ is not necessarily satisfied. When the antisymmetry law is not satisfied either, then the topologists speak of \emph{pseudo-quasi-metric spaces} \cite{Kim:pseudo-quasi-metric}. Lawvere \cite{LawvereFW:metric} observed that pseudo-quasi-metric spaces, which he called \emph{generalized metric spaces}, could be viewed as enriched categories \cite{KellyGM:book-enriched}. They are enriched over the additive monoid $[0,\infty]$, viewed as a monoidal category with a uniqe arrow $x\to y$ if and only if $x\geq y$. The distance $d(x,y) \in [0,\infty]$ is thus viewed as the 'hom-set' in the enriched sense. Lawvere's main result was the characterization of the Cauchy completion of a metric space as an enriched category construction. This view of distances and contractions turned out to provide an alternative to domains for denotational semantics \cite{WagnerK}, and their categorical completions were elaborated in \cite{RuttenJ:metric,Schellekens:Yoneda}. Distance spaces as defined in \ref{def-distance} are a special case of generalized metric spaces, since they are required to satisfy the antisymmetry law. This is mainly a matter of convenience, as the following lemma shows.

\begin{lemm} A map $d_A:A\times A\to [0,\infty]$ which is reflexive and transitive in the sense of Def.~\ref{def-distance} is also antisymmetric if and only if it satisfies either of the following equivalent conditions
\begin{itemize}
\item $\left(\forall z.\ \dis z x = \dis z y\right)\Rightarrow x=y$
\item $\left(\forall z.\ \dis x z = \dis y z\right)\Rightarrow x=y$
\end{itemize}
\end{lemm}

\bpr In the presence of transitivity and reflexivity, $\dis x y = 0$ holds if and only if $\forall z.\ \dis z x \geq \dis z y$, or equivalently if and only if 
$\forall z.\ \dis x z \leq \dis y z$. The result follows.
\epr

\begin{corr}
Distance spaces are just the \emph{skeletal}\/ generalized metric spaces.
\end{corr}

\subsection{Examples} 
The first example of a distance space is, of course, the interval $[0,\infty]$ itself, with the metric
\bea\label{vdash}
\dist {[0,\infty]} x y\ =\ x\multimap y  & = & \begin{cases} y-x & \mbox{ if } x\lt y\\
0 & \mbox{ otherwise}
\end{cases}
\eea
The $\multimap$ notation is convenient because the operation $d_{[0,\infty]} = \multimap : [0,\infty]\times [0,\infty]\to [0,\infty]$ makes $[0,\infty]$ into a closed category
\bea\label{MP}
x+y\  \geq \ z & \iff & x \geq y \multimap z
\eea
Any metric space is obviously an example of a distance space. But in distance spaces, the distance $d(a,b)$ from $a$ to $b$ does not have to be the same as the distance $d(b,a)$ from $b$ to $a$. E.g., $a$ may be on a hill, and $b$ in the valley, and traveling one way may be easier than traveling the other way. For our purposes described in the Introduction, this distinction is quite important, since a program constructing an attack $b$ from a system code $a$ does not have to be related in any obvious way to the program performing the construction the other way.

For a non-metric family of distance spaces, take any poset $(S, \underset S \sqsubseteq)$ and define a distance space $\left(\Forgg S, d_{\Forgg S}\right)$ by setting $\dist {\Forgg S} x y  = 0$ if $x\underset S \sqsubseteq y$, otherwise $\infty$. The other way around, any distance space $A$ induces two posets, $\Rightt A$ and $\Leftt A$, with the same underlying set and 
\[x\underset{\Rightt A}\sqsubseteq y \iff  \dist A x y = 0 \qquad \qquad \qquad x\underset{\Leftt A}\sqsubseteq y  \iff \dist A x y \lt \infty  
\]
The constructions $\Forgg$, $\Rightt$ and $\Leftt$ form the adjunctions $\Leftt \dashv \Forgg \dashv \Rightt : \Dist \to \Pos$. Since $\Forgg:\Pos \inclusion \Dist$ is an embedding, $\Pos$ is thus a reflective and correflective subcategory of $\Dist$. 

Distance spaces are thus a common generalization of posets and metric spaces. For an example not arising from posets or metric spaces, take any family of sets $\XXX \subseteq \WP X$, and define
\bea\label{terms}
d(x,y) & = & | y \setminus x |
\eea
The distance of $x$ and $y$ is thus the number of elements of $y$ that are not in $x$. If $X$ is a set of terms, say in a dictionary, and $\XXX$ is a set of documents, each viewed as a set of terms, then the distance between two documents is the number of terms that occur in one document and not in the other. In natural language processing, documents are usually presented as multisets (bags) of terms, and the distance is defined in terms of multiset subtraction, which generalizes the set difference used in \eqref{terms}. In any case, it is clear that the asymmetry of the notion of distance is as essential for such applications as it is for the one described in the Introduction. 

%
%
%

%
%
%
\subsection{Basic constructions}
Given two distance space $A$ and $B$, we define:
\begin{itemize}
\item \emph{dual} $\op A$: take the same underlying set and define the dual metric to be $\dist {\op A} x y = \dist A y x$;
\item \emph{product} $A\times B$: take the cartesian product of the underlying sets and set the product metric to be  $\dist{A\times B} {x, u} {y, v}= \dist A x y \vee \dist B u v$
\item the \emph{power} $B^A$: take the set of contractions $\Dist(A, B)$ to be the underlying set and set the metric to be $\dist {B^A} f g = \bigvee_{x\in A} \dist B {fx} {gx}$
\item the \emph{tensor} $A\otimes B$: take the cartesian product of the underlying sets and set the product metric to be  $\dist{A\otimes B} {x, u} {y, v}= \dist A x y + \dist B u v$.
\end{itemize}
These constructions induce the natural correspondences
\[\Dist(A, B)\times \Dist(A,C)  \cong  \Dist(A, B\times C)\quad\mbox{and} \quad 
\Dist(A\otimes B, C) \cong  \Dist(A, C^B)
\]

\paragraph{Terminology.} Contractions $f:A\to B$ are called \emph{covariant}, whereas contractions $f:\op A \to B$ are \emph{contravatiant}.

\section{Sequences and their limits}\label{sequences-sec}
\subsection{Left and right sequences}
Intuitively, to complete a metric space means to add enough points so that every suitably convergent sequence has a limit. But usually many different sequences have the same limit. The main problem of the standard theory of completions is to recognize such sequences. The categorical approach overcomes this problem by considering \emph{canonical}\/ sequences. Instead of the sequences $s,t:\NNn\to A$ such that $\lim_{i\to \infty} s_i = \psi = \lim_{i\to \infty} t_i$, we consider a canonical sequence $\psi : A \to [0,\infty]$ where $\psi x $ intuitively denotes the distance from $\psi$ to $x$.

\begin{defn} \label{infsup}  In a distance space $A$, a \emph{(canonical) sequence}  is defined to be a contraction into $[0,\infty]$. More precisely, we define that
\begin{itemize}
\item a \emph{left} sequence is a covariant contraction  $\up \lambda : A\to [0,\infty]$\\
-- we write its value at $x\in A$ as $\up \lambda  x$
\medskip
\item a \emph{right} sequence is a contravariant contraction $\dow \varrho: \op A \to [0,\infty]$\\
-- we write its value at $x\in A$ as $x \dow \varrho $.
\end{itemize}
Each of the sets of sequences
\[
\Up A = \op{\left([0,\infty]^{A}\right)} \qquad \mbox{ and} \qquad \Do A = [0,\infty]^{\left(\op A\right)}
\]
forms a distance space, with the metrics
\[
\dist {\Up A} {\up \lambda} {\up \theta} = \bigvee_{x\in A} \up \theta x\multimap  \up \lambda x \qquad \mbox{ and} \qquad \dist {\Do A} {\dow \varrho} {\dow \mu } = \bigvee_{x\in A} x \dow \varrho  \multimap x \dow \mu 
\]
\end{defn}

\paragraph{Remarks.} The conditions $\dist A x y \geq  \up \lambda x\multimap  \up \lambda y$ and $\dist A x y \geq y \dow \varrho \multimap x \dow \varrho $, which say that $\up \lambda$ and $\dow \varrho$ are left and right contraction respectively, are by \eqref{MP} respectively equivalent to 
\[
 \up \lambda x + \dis x y \geq \up \lambda y \qquad \qquad \dis x y + y \dow \varrho \geq x \dow \varrho
\]

\subsection{Limits}\label{Limits-sec}
\begin{defn}
An element $u$ of a distance space $A$ is an\/ \emph{upper bound} of a right sequence $\dow \varrho$ in $A$ if for all $x\in A$ holds
\bea
x \dow \varrho &\geq & \dist A x u
\eea

An element $\ell$ of a distance space $A$ is a\/ \emph{lower bound} of a left sequence $\up \lambda$ in $A$ if for all $y\in A$ holds
\bea
\up \lambda y  &\geq & \dist A \ell y
\eea
\end{defn}

\begin{proposition}
An element $u\in A$ is an upper bound $\dow \varrho$ and $\ell\in A$ is a lower bound of $\up \lambda$ if and only if the following conditions hold for all $x, y \in A$
\bea
\dist A u y & \geq & \bigvee_{x\in A} x \dow \varrho \multimap \dist A x y \label{upperr}
\\
\dist {A} {x} {\ell} & \geq & \bigvee_{y\in A} \up \lambda y \multimap \dist A x y \label{lowerr}
\eea
\end{proposition}

\bpr
Condition \eqref{MP} implies that \eqref{upper} and \eqref{lower} are respectively equivalent with
\bea
x\dow \varrho + \dist A u y & \geq & \dist A x y \label{upper}\\
\dist {A} {x} {\ell} + \up \lambda y & \geq & \dist A x y \label{lower}
\eea
The claim follows by instantiating $y$ to $u$ in \eqref{upperr} and $x$ to $\ell$ in \eqref{lowerr}.
\epr

\begin{defn}
The \emph{supremum} $\tcoprod \dow \varrho$ of the right sequence $\dow \varrho$ and the \emph{infimum} $\tprod \up\lambda$ of the left sequence $\up \lambda$ are the elements of $A$ that satisfy for every $x,y\in A$
\bea
\dist A {\tcoprod \dow \varrho} y & = & \bigvee_{x\in A} x \dow \varrho\multimap  \dist A x y\label{pjoin}\\
\dist A x {\tprod \up \lambda}  & = & \bigvee_{y\in A} \up \lambda y \multimap \dist A x y\label{pmeet}
\eea
Suprema and infima constitute the \emph{limits} of a distance space.

The distance space $A$ is right (resp. left) \emph{complete} if every right (resp. left) sequence has a limit. The suprema and the infima thus yield the operations 
\[\tcoprod:\ \ \Do A \to A \qquad \mbox{and} \qquad \tprod  :\ \ \Up A \to A\]
\end{defn}

One apparent shortcoming of treating sequences categorically, i.e. saturating them to canonical sequences, is that it is not obvious how to define continuity, i.e. how to distinguish the contractions which preserve suprema or infima. Clearly, a left continuous contraction $f:A\to B$ should map the infimum of a left sequence $\up \lambda$ in $A$ into the infimum of the $f$-image of $\up \lambda$ in $B$. But what is the $f$-image of $\up \lambda : A \to [0,\infty]$ in $B$? This question calls for a slight generalization of the concept of sequence, and limit.

\subsection{Weighted limits} 
Limits are a special case of \emph{weighted limits}, which are studied in general enriched categories \cite[Ch.~3]{KellyGM:book-enriched}.  We just sketch theory of weighted limits in distance spaces.

\begin{defn}
For distance spaces $A$ and $K$  we define
\begin{itemize}
\item \emph{left diagrams} as pairs of contractions $\left<k:K\to A,\up \lambda:K\to [0,\infty]\right>$ 
\item \emph{right diagrams} as pairs of contractions $\left<k:K\to A,\dow \varrho : \op K\to [0,\infty]\right>$ 
\end{itemize}
\end{defn}

\paragraph{Terminology and notation.} The component $k:K\to A$ of a diagram is called its \emph{shape}. Using the angular brackets to denote the functions into cartesian products, we also write
\begin{itemize}
\item $\left<k,\up \lambda\right>:K\to A\times  [0,\infty]$ for $\left<k:K\to A,\up \lambda:K\to [0,\infty]\right>$
\item $\left<k,\op{\dow \varrho}\right> : K\to A\times  \op{[0,\infty]}$ for $\left<k:K\to A,\dow \varrho : \op K\to [0,\infty]\right>$ 
\end{itemize}

\begin{defn}
The \emph{weighted supremum} $\tcoprod_{\dow \varrho} k$ of the right diagram $<k,\op{\dow \varrho}>: K\to A \times \op{[0,\infty]}$ and the \emph{weighted infimum} $\tprod_{\up\lambda} k$ of the left diagram $<k,\up\lambda>: K\to A \times [0,\infty]$ 
are the elements of $A$ that satisfy for every $x,y\in A$
\bea
\dist A {\tcoprod_{\dow \varrho} k} y & = & \bigvee_{x\in K} x \dow \varrho \multimap \dist A {kx}{y}\label{wjoin}\\
\dist A x {\tprod_{\up\lambda} k}  & = & \bigvee_{y\in K} \up \lambda y \multimap \dist A {x} {k{y}} \label{wmeet}
\eea
\end{defn}

\paragraph{Remarks.} Limits arise as a special case of weighted limits, by viewing sequences as diagrams of shape $k = \id:A\to A$. A contraction $f:A\to B$ thus maps, say, a left sequence $<\id,\up \lambda> : A\to A\times [0,\infty]$ to the diagram $<f ,\up \lambda> : A\to B\times [0,\infty]$ in $B$. More generally, it maps a left sequence $<k,\up \lambda> : K\to A\times [0,\infty]$ to the diagram $<f \circ k,\up \lambda> : K\to B\times [0,\infty]$ in $B$. It is thus clear and easy to state what it means that a contraction preserves a weighted limit.

\be{defn}A contraction $f:A\to B$ preserves 
\begin{itemize}
\item weighted suprema if  $f\left(\tcoprod_{\dow \varrho} k\right) \ = \ \tcoprod_{\dow \varrho} (f\circ k)$, and
\item weighted infima if $f\left(\tprod_{\up\lambda} k\right) \ = \ \tprod_{\up\lambda} (f\circ k)$.
\end{itemize}
\ee{defn}

On the other hand, although convenient to work with, weighted limits of diagrams in distance spaces also boil down to the limits of suitable sequences. We just state this fact, since it simplifies the construction of the completions; but leave the proof for another paper, since the proof construction is not essential for the goal of the present paper.

\be{proposition}
A distance space has 
\begin{itemize}
\item the weighted suprema of all right diagrams if and only if it has the suprema of all right sequences;
\item the weighted infima of all left diagrams if and only if it has the infima of all left sequences.
\end{itemize}

\ee{proposition}

\subsection{Completions}
Every element $a$ of a distance space $A$ induces two \emph{representable} sequences
\begin{alignat*}{4}
{\rm \cmn} a\ :\ A&\to  [0,\infty] & \qquad \qquad && \mnd a\ :\ \op A&\to  [0,\infty]\\
x & \mapsto \dist A a x  & && x &\mapsto \dist A x a
\end{alignat*}
These induced contractions $\cmn : A\to \Up A$ and $\mnd : A \to \Do A$ correspond to the \emph{Yoneda-Cayley embeddings} \cite[Sec.~III.2]{MacLane:CWM}. They make $\Up A$ into the lower completion, and $\Do A$ into the upper completion of the distance space $A$. 
\begin{proposition}\label{completion}
$\Up A$ is left complete and $\Do A$ is right complete. Each of them is universal among distance spaces with the corresponding completeness properties, in the sense that
\begin{itemize}
\item any monotone $f:A\to C$ into a complete distance space $C$ induces a unique $\prod$-preserving morphism $f_\#: \Up A\to C$ such that $f = f_\#\circ \cmn$;
\item any monotone $g:A\to D$ into a cocomplete distance space $D$ induces a unique $\coprod$-preserving morphism $g^\#: \Do A\to D$ such that $g = g^\#\circ \mnd$.
\end{itemize}
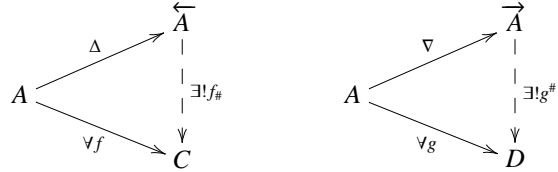
\[
\xymatrix@-1pc{
&&& \Up A \ar@{-->}[dd]^{\exists ! f_\#}\\
A\ar[urrr]^{\cmn} \ar[drrr]_{\forall f} \\
&&& C
}\qquad \qquad 
\xymatrix@-1pc{
&&& \Do A \ar@{-->}[dd]^{\exists ! g^\#}\\
A\ar[urrr]^{\mnd} \ar[drrr]_{\forall g} \\
&&& D
}
\] 
\end{proposition}

These constructions for have been thoroughly analyzed in \cite{RuttenJ:metric,Schellekens:Yoneda}. Here we just state the basic facts that justify our notations, and substantiate the further developments.

\begin{proposition}\label{yoneda} ("The Yoneda Lemma") For every ${\dow {\varrho}}\in \Do A$ and ${\up {\lambda}} \in \Up A$ and  holds
\bear
a \dow \varrho  & = & \bigvee_{x\in A}  x \left(\mnd a\right) \multimap x\dow \varrho\ =\ \dist {\Do A} {\mnd a} {\dow \varrho}
\\
 \up \lambda b  & = &
\bigvee_{x\in A}   \left(\cmn b\right) x \multimap {\up \lambda}x \ =\ \dist {\Up A} { \up \lambda } {\cmn b} 
\eear
\end{proposition}

Instantiating in the preceding proposition $\up \lambda$ to $\cmn a$ and $\dow \varrho$ to $\mnd b$ yields
\begin{corr}
The embeddings $\cmn : A\to \Up A$ and $\mnd : A \to \Do A$ are isometries
\[
\dist A a b\  =\  \dist {\Do A} {\mnd a} {\mnd b}\ =\ \dist {\Up A} { \cmn a} {\cmn b}  \]
\end{corr}

\subsection{Adjunctions}
\paragraph{Notation.} In any distance space $A$, if is often convenient to abbreviate $\dist A x y = 0$ to $x \underset A \closeto y$. For $f,g:A\to B$, it is easy to see that $f\underset{B^A} \closeto g$ if and only if $fx\underset B \closeto gx$ for all $x\in A$.

\begin{proposition}\label{adjunctions}
For any contraction $f:A\to B$ holds 
\[ (a)\iff(b)\iff(c) \quad\mbox{ and }\quad (d)\iff(e)\iff(f)\] 
where
\begin{anumerate}
\item $f\left(\tcoprod \dow \varrho\right) = \tcoprod_f\left(\dow \varrho\right)$
\item $\exists f_\ast:B\rightarrow A\ \forall x\in A\ \forall y\in B.\  \dist B {fx} y = \dist A x {f_\ast y}$
\item $\exists f_\ast:B\rightarrow A.\  \id_A \closeto {f_\ast f} \wedge {ff_\ast} \closeto \id_B$
\item $f\left(\tprod \up \lambda\right) = \tprod_f\left(\up \lambda\right)$
\item $\exists f^\ast : B\rightarrow A\ \forall x\in A\ \forall y\in B.\  \dist B {f^\ast y} x = \dist A y {fx}$
\item $\exists f^\ast:B\rightarrow A.\  {f^\ast f} \closeto \id_A \wedge \id_B \closeto {ff^\ast}$
\end{anumerate}
Each of the morphisms $f^\ast$ and $f_\ast$ is uniquely determined by $f$, whenever they exist.
\end{proposition}

\begin{defn}
A \emph{right adjoint} is a contraction satisfying (a-c) of Prop.~\ref{adjunctions}; a \emph{left adjoint} satisfies (d-f).
A \emph{(distance) adjunction} between the distance spaces $A$ and $B$ is a pair of contractions $f^\ast: A\rightleftarrows B: f_\ast$ related as in (b-c) and (e-f). 
\end{defn}

Since $f\closeto g$ and $g\closeto f$ implies $f = g$, it follows that

\begin{corr}\label{Corr:adjunctions}
For every adjunction $f^\ast: A\rightleftarrows B: f_\ast$ holds
\[
f^\ast f_\ast f^\ast = f^\ast \qquad \qquad f_\ast f^\ast f_\ast = f_\ast
\]
\end{corr}

Equations \eqref{pjoin} and \eqref{pmeet} immediately yield the following fact.

\begin{proposition} \label{limits-adjoints}
Limits are adjoints to the Yoneda-Cayley embeddings:
\[
\dist A {\tcoprod \dow \varrho} y = \dist {\Do A} {\dow \varrho} {\mnd y} \quad\mbox{ and } \quad
\dist A x {\tprod \up \lambda}   = \dist {\Up A} {\cmn x} {\up \lambda} 
\]
\end{proposition}

Putting  Propositions \ref{adjunctions} and \ref{limits-adjoints} together yields yet another familiar fact.

\be{proposition} The sup-completion $\mnd:A\to \Do A$ preserves any infima that exist in $A$. The inf-completion $\cmn:A\to \Up A$ suprema that exist exist in $A$. 
\ee{proposition}

%
%
%
%
%

\subsection{Projectors and nuclei}
\begin{proposition}\label{projections}
For any adjunction $f^\ast: A\rightleftarrows B: f_\ast$ holds 
\[ (a)\iff(b) \quad\mbox{ and }\quad (c)\iff(d)\] 
where
\begin{anumerate}
\item $\forall xy\in B.\ \dist A {f_\ast x} {f_\ast y} = \dist B x y$
\item $f^\ast f_\ast = \id_B$
\item $\forall xy\in A.\ \dist B {f^\ast x} {f^\ast y} = \dist A x y$
\item $f_\ast f^\ast  = \id_A$
\end{anumerate}
\end{proposition}

\begin{defn}
A map $g$ from a distance space $A$ to a distance space $B$ is an \emph{embedding} if it preserves the distance, i.e. satisfies $\dist A x y = \dist B {gx} {gy}$ for all $x,y\in A$. An adjoint of an embedding is called a \emph{projection}.

An adjunction $p^\ast: A\rightleftarrows B: e_\ast$ of a left projection and right adjoint, as in Prop.~\ref{projections}(a-b),  is called a \emph{reflection}. An adjunction $e^\ast: A\rightleftarrows B: p_\ast$ of a left embedding and right projection, as in Prop.~\ref{projections}(c-d), is called a \emph{coreflection}. 
\end{defn} 

\begin{defn}
A \emph{nucleus} of the adjunction $f^\ast: A\rightleftarrows B: f_\ast$ consists of a distance space $\cuu f$ together with
\begin{itemize}
\item embeddings $A \stackrel {e_\ast} \clusionin \cuu f \stackrel{e^\ast} \inclusion B$
\item projections $A \stackrel {p^\ast} \epi \cuu f \stackrel{p_\ast} \ipe B$
\end{itemize}
such that $f^\ast = e^\ast p^\ast$ and $f_\ast = e_\ast p_\ast$. 
\end{defn}

\begin{proposition}\label{factorization}
Any adjunction factors through its nucleus by reflection followed by a coreflection. The nucleus of the adjunction $f^\ast: A\rightleftarrows B: f_\ast$ is in the form
\bea\label{eq-nucleus}
\cuu f & = & \left\{<x,y>\in A\times B\ |\  f^\ast x = y \wedge x = f_\ast y\right\}
\eea
and the factoring is
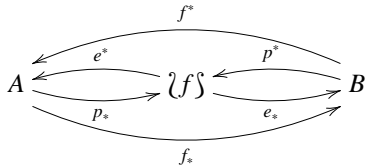
\[\xymatrix@C+2pc{
A \ar@/_/[r]_{p_\ast} \ar@<-1.2ex>@/_1.3pc/[rr]_{f_\ast} &\cuu f \ar@/_/[r]_{e_\ast} \ar@/_/[l]_{e^\ast}& B  \ar@/_/[l]_{p^\ast}  \ar@<-1.2ex>@/_1.3pc/[ll]_{f^\ast}
}\]
Any right adjoint factors through the nucleus by a right projection followed by a right embedding, and any left adjoint factors through the nucleus by a left projection followed by a left embedding. This factorization is unique up to isomorphism.
\end{proposition}

\bpr
For any adjunction $f^\ast: A\rightleftarrows B: f_\ast$, form the distance spaces 
\begin{gather*}
\cuu f_A = \{x\in A\ |\  f_\ast f^\ast x = x\}\qquad \qquad \cuu f _B = \{y\in B\ |\  f^\ast f_\ast y = y\}
\end{gather*}
are easily seen to be isomorphic with the nucleus. The factorisation is thus
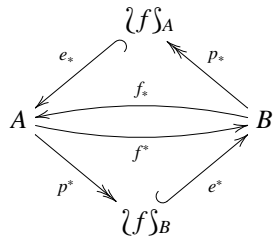
\[\xymatrix{
& \hspace{.7em} \cuu f_A \ar@{^{(}->}[dl]_{e_\ast} \\
A \ar@/_/[rr]_{f^\ast} \ar@{->>}[dr]_{p^\ast} && B  \ar@/_/[ll]_{f_\ast} \ar@{->>}[ul]_{p_\ast} \\
& \hspace{.4em} \cuu f_B \ar@{^{(}->}[ur]_{e^\ast}
}\]
\epr

\subsection{Cones and cuts}
The \emph{cone extensions} are the contractions $\cmn^\#$ and $\mnd_\#$
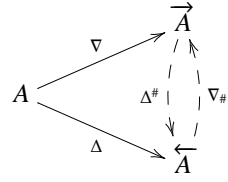
\begin{gather*}
\xymatrix@-1pc{
&&& \Do A \ar@/_/@{-->}[dd]_{\cmn^\#}\\
A\ar[urrr]^{\mnd} \ar[drrr]_{\cmn} \\
&&& \Up A \ar@/_/@{-->}[uu]_{\mnd_\#}
}\\
a \left(\cmn^\# \dow \varrho\right) = \bigvee_{x\in A} x \dow \varrho \multimap \dis x a \qquad \qquad \left(\mnd_\# \up \lambda \right) a = \bigvee_{x\in A} \up \lambda x \multimap \dis a x
\end{gather*} 
induced by the universal properties of the Yoneda embeddings $\mnd$ and  $\cmn$, as per Prop.~\ref{completion}. Since $\cmn^\#$ thus preserves suprema, and $\mnd_\#$ preserves infima, Prop.~\ref{adjunctions} implies that each of them is an adjoint, and it is not hard to see that they are adjoint to each other. 

\begin{lemma} \label{Lemma:adjunction}
The adjunction $\cmn^\# : \Do A \rightleftarrows \Up A: \mnd_\#$ is witnessed by 
\[
\dow \varrho \closeto \mnd_\# \cmn^\# \dow\varrho
\qquad\qquad
\up \lambda   \closeto \cmn^\# \mnd_\# \up \lambda
\]
\end{lemma}

\bpr
Unfolding the definitions of $\mnd_\#$ and $\cmn^\#$ gives
\bear
a \left(\mnd_\# \cmn^\# \dow\varrho\right) & = & \bigvee_{u\in A} \left(\bigvee_{x\in A} x \dow \varrho \multimap \dis x u\right) \multimap \dis a u
\eear
which shows that the first claim follows from the fact that for every $u\in A$ holds
\[\prooftree
\bigvee_{x\in A} x \dow \varrho \multimap \dis x u\  \geq\ a \dow \varrho \multimap \dist A a u \justifies
\prooftree
a \dow \varrho + \left(\bigvee_{x\in A} x \dow \varrho \multimap \dis x u\right)\  \geq\  \dist A a u 
\justifies
a \dow \varrho   \geq \left(\bigvee_{x\in A} x\dow \varrho \multimap \dis x u\right) \multimap \dist A a u 
\endprooftree
\endprooftree\]
\epr

\begin{proposition}\label{saturated-prop}
For every $\dow \varrho \in \Do A$ every $\up \lambda \in \Up A$ holds
\begin{alignat}{5}
\dow \varrho  & = \mnd_\# \cmn^\# \dow\varrho
&\quad\iff \quad& 
\mnd_\# \cmn^\# \dow\varrho \closeto \dow \varrho 
&\quad \iff\quad  \exists \up\lambda.\ \dow \varrho = \mnd_\#\up \lambda \label{eq:firstl}\\
\up \lambda  & = \cmn^\# \mnd_\# \up \lambda
&\quad\iff \quad&
\cmn^\# \mnd_\#  \up \lambda \closeto \up \lambda 
&\quad \iff\quad \exists \dow \varrho.\ \up \lambda = \cmn^\#\dow \varrho\label{eq:secondl}
\end{alignat}
\end{proposition}

\bpr
The first equivalence in each \eqref{eq:firstl} and \eqref{eq:secondl} follows from Lemma~\ref{Lemma:adjunction}. The second two equivalences follow from $\cmn^\# \mnd_\#\cmn^\# = \cmn^\#$ and $\mnd_\#\cmn^\# \mnd_\#= \mnd_\#$, which is the property of all distance adjunctions, by Corollary \ref{Corr:adjunctions}.
\epr

\begin{corr}
The following distance spaces isomorphic:
\bear
\left(\Do A\right)_{\mnd_\# \cmn^\#} & = & \left\{\dow \varrho\in \Do A\ |\ \dow \varrho  = \mnd_\# \cmn^\# \dow\varrho\right\}\\
\left(\Up A\right)_{\cmn^\#\mnd_\# } & = & \left\{\up\lambda \in \Up A\ |\ \up \lambda = \cmn^\# \mnd_\#  \up \lambda\right\}
\eear
\end{corr}


\begin{defn}
The \emph{cones} in a distance space $A$ are the sequences in $\left(\Do A\right)_{\mnd_\# \cmn^\#}$ and $\left(\Up A\right)_{\cmn^\#\mnd_\# }$.  A \emph{cut} in $A$ is  a pair of cones $\gamma = <\dow \gamma, \up \gamma> \in \left(\Do A\right)_{\mnd_\# \cmn^\#}\times  \left(\Up A\right)_{\cmn^\#\mnd_\# }$ such that $\dow \gamma = \mnd_\# \up \gamma$. The set of cuts is denoted by $\UD A$.
\end{defn}

\begin{lemm}
There are bijections $\left(\Do A\right)_{\mnd_\# \cmn^\#} \cong \UD A \cong \left(\Up A\right)_{\cmn^\#\mnd_\# }$, extending the isomorphism $\left(\Do A\right)_{\mnd_\# \cmn^\#} \cong \left(\Up A\right)_{\cmn^\#\mnd_\# }$ from Prop.~\ref{saturated-prop}.
\end{lemm}

\begin{proposition}\label{ud-cocompl}
The set of cuts $\UD A$ with the distance defined by
\[
\dist {\UD A} \gamma \varphi \ =\ \dist {\Do A} {\dow \gamma} {\dow \varphi} \ =\ \dist {\Up A} {\up \gamma} {\up \varphi}
\]
is a left and right complete distance space. 
\end{proposition}

\paragraph{Notation.} We often abuse notation and write
\begin{itemize}
\item $\up \varrho$ for the associated cone $\mnd_\#\dow \varrho$, and
\item $\dow \lambda$ for the associated cone $\cmn^\#\up \lambda$.
\end{itemize}

\bprf{ of Prop.~\ref{ud-cocompl}}
The $\UD A$-infima are constructed in $\Do A$, the $\UD A$-suprema in $\Up A$. To spell this out, consider $\up \lambda : \UD A \to [0,\infty]$ and $\dow \varrho: \op{\UD A} \to [0,\infty]$. Extend them along the isomorphisms 
$$\left(\Do A\right)_{\mnd_\# \cmn^\#}\ \cong\ \UD A\cong \left(\Up A\right)_{\cmn^\#\mnd_\# }\ \cong\  \UD A$$ 
to get 
$\up \lambda : \left(\Do A\right)_{\mnd_\# \cmn^\#} \to [0,\infty]$ and $\dow \varrho: \left(\Up A\right)^o_{\cmn^\#\mnd_\# } \to [0,\infty]$. Then
\[
\prod \up \lambda\  =\  \up \lambda \circ \mnd\   \in\  \left(\Do A\right)_{\mnd_\# \cmn^\#}\qquad   \qquad \qquad \coprod \dow \varrho\  =\  \dow \varrho \circ \cmn\   \in\  \left(\Up A\right)_{\cmn^\#\mnd_\# }
\]
The claim now boils down to showing that  the inclusion $ \left(\Do A\right)_{\mnd_\# \cmn^\#}\inclusion \Do A$ preserves infima, whereas the inclusion $\left(\Up A\right)_{\cmn^\#\mnd_\# }\inclusion \Up A$ preserves the suprema. But this is immediate from the next Lemma.
\epr

\begin{lemm}\label{limits-presheaves} The limits of the cut sequences
\begin{alignat*}{3}
\dow \Upsilon & : \op{\Do A} \to [0,\infty] & \qquad\qquad \qquad&  \up \Lambda & : \Do A \to [0,\infty]\\
\dow K & : \op{\Up A} \to [0,\infty] & & \up \Psi & : \Up A \to [0,\infty]
\end{alignat*}
can be computed as follows
\begin{alignat*}{3}
a\left(\tcoprod \dow \Upsilon  \right) \ & = \bigwedge_{\dow \xi\in \Do A} a \dow \xi + \dow \xi \dow \Upsilon  &\qquad\qquad &\qquad &   \left(\tprod \up \Lambda\right) a\  & =\ 
\up \Lambda ({\mnd a})
\\
a \left(\tcoprod \dow K\right) & =\ 
({\cmn a})\dow K
& & & \left(\tprod \up \Psi  \right) a & = \bigwedge_{\up \zeta\in \Up A}  
 \up \Psi {\up \zeta} +  \up \zeta a 
\end{alignat*}
\end{lemm}

\begin{corr}\label{sup-inf}
A distance space $A$ has all suprema if and only if it has all infima. 
\end{corr}

%

\paragraph{Dedekind-MacNeille completion is a special case.} If $A$ is a poset, viewed as the distance space $\Forgg A$, then $\UD \Forgg A$ is the Dedekind-MacNeille completion of $A$. The above construction extends the Dedekind-MacNeille completion of posets \cite{MacNeille} to distance spaces, in the sense that it satisfies in the same universal property, spelled out in \cite{Banaschewski-Bruns}.

\section{Distance matrices}\label{Matrices-sec}

\subsection{Definitions}
\begin{defn}\label{matrix-def}
A {\em distance matrix} $\Phi$ from distance space $A$ to distance space $B$ is a sequence $\Phi: \op A \times B\to [0,\infty]$. We denote it by $\Phi:A\bito B$, and the value of $\Phi$ at $x\in A$ and $y\in B$ is written $\mol \Phi x y$. The matrix composition of $\Phi:A\bito B$ and $\Psi: B\bito C$ is defined
 \bear
 \mol{(\Phi\, ; \Psi)} x z & = & \bigwedge_{y\in B} \mol \Phi x y + \mol \Psi y z 
 \eear 
With this composition and the identities $\Id_A:A\bito A$ where $x(\Id_A)x' \ = \ \dist A x {x'}$, distance spaces and distance space matrices form the category $\Matr$.
 \end{defn}
 
\paragraph{Remark.} Note that the defining condition $\dist A u x + \dist B y v \geq \dis {\mol \Phi x y} {\mol \Phi u v}$, which says that $\Phi$ is a contraction $\op A \times B\to [0,\infty]$, can be equivalently written
\bea\label{matrix-cond}
\dist A u x + \mol \Phi x y + \dist B y v \ \geq \ \mol \Phi u v
\eea

\begin{defn} 
Transposing the indices yields the \emph{transposed} matrix: 
\[
\prooftree
\Phi\  :\  A\bito B\ :\  \mol \Phi x y 
\justifies
\op\Phi\  :\  \op B\bito \op A\ :\  \mol {\op\Phi} y x 
\endprooftree
\]
The \emph{dual} $\dual \Phi:B\bito A$ of a matrix $\Phi:A\bito B$ has the entries
\[
\prooftree
\Phi\  :\  A\bito B\ :\  \mol \Phi x y \hspace{14em}
\justifies
\dual \Phi\  :\  B\bito A\ :\  \mol {\dual \Phi} y x = \bigvee_{\substack{u\in A\\ v\in B}} \mol \Phi u v \multimap \left( \dist A u x +  \dist B y v \right)
\endprooftree
\]
A matrix $\Phi:A\bito B$ where $\Phi^{\ddag\ddag} = \Phi$ is called a \emph{suspension}.
\end{defn}

\paragraph{Remarks.} 
The transposition is obviously an involutive operation, i.e. $\Phi^{oo} = \Phi$. 
It is easy to derive from Prop.~\ref{saturated-prop} that $\dist \Phi x y \geq \dist {\Phi^{\ddag\ddag}} x y$ holds for all $x\in A$ and $y\in B$, and that $\Phi = \Phi^{\ddag\ddag}$ holds
 if and only if there is some $\Psi: B\bito A$ such that $\Phi = \Psi^\ddag$. Since $\Phi \closeto \Psi\Rightarrow\dual \Psi \closeto \dual \Phi$, it follows that  $\Phi \closeto \Phi^{\ddag\ddag}$ implies $\dual \Phi = \Phi^{\ddag\ddag\ddag}$.


\begin{proposition}\label{Galois-prop}
$\Phi:A\bito B$ and $\Phi^\ddag:B\bito A$  satisfy $\Phi\, ;\Phi^\ddag \closeto \Id_A$ and $\Phi^\ddag\, ;\Phi \closeto \Id_B$.
\end{proposition}

\bpr  The condition $\Phi\, ;\dual \Phi \closeto \Id_A$ is proven as follows:
\[\prooftree
\prooftree
\bigvee_{\substack{u\in A\\ v\in B}} \mol \Phi u v \multimap \left( \dist A  u {x'} + \dist B y v\right)\ \  \geq\ \  \mol {\Phi} x y \multimap \dist A x {x'} 
\justifies
\mol {\Phi} x y + \left(\bigvee_{\substack{u\in A\\ v\in B}} \mol \Phi u v \multimap \left( \dist A u {x'} + \dist B y v\right) \right)\ \  \geq\ \   \dist A x {x'}
\endprooftree
\justifies
\mol {\Phi} x y + \mol {\dual \Phi} y {x'}\ \  \geq\ \   \dist A x {x'}
\endprooftree\]
The second condition is proven analogously.
\epr


\begin{defn}\label{density-def}
A matrix $\Phi: A\bito B$ is  \emph{embedding} if $\Phi\, ; \dual \Phi = \Id_A$; and a \emph{projection} if $\dual \Phi\, ; \Phi = \Id_B$.
\end{defn}

\begin{defn}\label{decomposition-def}
A \emph{decomposition} of a matrix $\Phi:A\bito B$ consists of a distance space $D$, with
\begin{itemize}
\item projection matrix $P: A\bito D$, i.e. $\dist D d {d'} = \bigwedge_{x\in A} \mol {\dual P} d x + \mol P x {d'}$,
\item embedding matrix $E: D\bito B$, i.e. $\dist D d {d'}  =  \bigwedge_{y\in B} \mol {E} d y + \mol {\dual E} y {d'}$,
\end{itemize}
such that $\Phi = P\, ; E$, i.e. $\mol \Phi x y  =  \bigwedge_{d\in D} \mol P x d + \mol E d y$.\end{defn}

%
%
%
%
%
%
%

\paragraph{Matrices as adjunctions.} A matrix $\Phi : A\bito B$ can be equivalently presented as either of the two contractions $\Phi_\bullet$ and $\Phi^\bullet$, which extend to $\Phi_\ast$ and $\Phi^\ast$ using Prop.~\ref{completion}
\[
\prooftree
\prooftree
\op A\times B\tto \Phi [0,\infty]
\justifies
A \tto {\Phi_\bullet} \Up B\qquad \qquad B \tto {\Phi^\bullet} \Do A
\endprooftree
\justifies
\Do A \tto {\Phi^\ast} \Up B\qquad \qquad \Up B \tto {\Phi_\ast} \Do A
\endprooftree
\] 
\vspace{-1\baselineskip}
\bea\label{adj}
\left(\Phi^\ast\dow \varrho\right)b  =  \bigvee_{x\in A} x \dow \varrho \multimap \mol \Phi x b\qquad\qquad
\left(\Phi_\ast\up \lambda\right)a =  \bigvee_{y\in B} \up \lambda y \multimap \mol \Phi a y
\eea
Both extensions, and their nucleus, are summarized in the following diagram
 \begin{equation}\label{decomp}
 \begin{split}
\xymatrix
{
A \ar[rrr]^{\mnd} \ar@{o.>}[dd]|\Phi \ar[ddrrr]^(.3){\Phi^\bullet} &&& \Do A \ar@/_/@{-->}[dd]_{\Phi^\ast}
\\ &&& 
&& 
\ \ \cuu \Phi 
\ar@/_/@{^{(}->}[ull]_{e_\ast} 
\ar@/_/@{^{(}->}[dll]_{e^\ast}
\ar@/^/@{<<-}[ull]_{p^\ast} 
\ar@/^/@{<<-}[dll]_{p_\ast}
\\
B \ar[rrr]_\cmn \ar[uurrr]_(.3){\Phi_\bullet}&&& \Up B \ar@/_/@{-->}[uu]_{\Phi_\ast}
}
\end{split}
\end{equation}
%
The adjunction $\Phi^\ast:\Do A\rightleftarrows \Up B:\Phi_\ast$ means that
\[
\dist {\Up B} {\Phi^\ast \dow \varrho} {\up \lambda}\ =\ \bigvee_{y\in B}  {\up \lambda} y \multimap (\Phi^\ast \dow \varrho) y \ = \ \ \bigvee_{x\in A}  x{\dow \varrho} \multimap x (\Phi_\ast \up \lambda)\ =\  \dist {\Do A} {\dow \varrho} {\Phi_\ast \up \lambda}
\]
holds. The other way around, it can be shown that any adjunction between $\Do A$ and $\Up B$ is completely determined by the induced matrix from $A$ to $B$. 

\begin{proposition}
The matrices $\Phi\in \Matr(A,B)$ are in a bijective correspondence with the adjunctions $\Phi^\ast: \Do A\rightleftarrows \Up B : \Phi_\ast$.
\end{proposition}

\begin{lemm}\label{repres-matrix-lemma}
 $\dist {\Up B} {\Phi^\ast \mnd x} {\cmn y}\  =\  \mol \Phi x y\  =\  \dist{\Do A} {\mnd x} {\Phi_\ast \cmn y}$
\end{lemm}

\subsection{Decomposition through nucleus}
%
\begin{proposition}\label{matrix-saturated-prop}
For every $\dow \alpha \in \Do A$ every $\up \beta \in \Up B$ holds
\begin{alignat*}{5}
\dow \alpha  
& \closeto \Phi_\ast \Phi^\ast \dow\alpha
&\qquad\mbox{and}\qquad& 
\Phi_\ast \Phi^\ast \dow\alpha\closeto \dow \alpha 
&\ \iff \exists \up\beta\in \Up B.\ \dow \alpha = \Phi^\ast\up \beta\\
\up \beta  & \closeto \Phi^\ast \Phi_\ast \up \beta
&\qquad\mbox{and}\qquad&
\Phi^\ast \Phi_\ast  \up \beta\closeto \up \beta
&\ \iff \exists \dow \alpha \in \Do A.\ \up \beta = \Phi_\ast\dow \alpha
\end{alignat*}
The adjunction $\Phi^\ast:A\rightleftarrows B:\Phi_\ast$ induces the isomorphisms between the following distance spaces
\bear
\cuu \Phi_A & = & \left\{\dow \alpha\in \Do A\ |\ \dow \alpha  = \Phi_\ast \Phi^\ast \dow\alpha\right\}\\
\cuu \Phi_B & = & \left\{\up\beta \in \Up B\ |\ \up \beta = \Phi^\ast \Phi_\ast  \up \beta\right\}\\
\cuu \Phi & = & \left\{\gamma = <\dow \gamma, \up \gamma> \in \Do A\ \times\! \Up B\ |\ \dow \gamma  = \Phi_\ast \up \gamma \wedge \Phi^\ast \dow \gamma = \up\gamma\right\}
\eear
with the metric
\[
\dist {\cuu\Phi} \gamma \varphi \ =\ \dist {\Do A} {\dow \gamma} {\dow \varphi} \ =\ \dist {\Up B} {\up \gamma} {\up \varphi}
\]
\end{proposition}

\begin{defn}
$\cuu \Phi$ is called the \emph{nucleus} of the matrix $\Phi$. Its elements are the \emph{$\Phi$-cuts}. 
\end{defn}

\begin{thrm}\label{complete-thm}
The nucleus $\cuu \Phi$ 
of the adjunction $\Phi^\ast : \Do A \rightleftarrows \Up B:\Phi_\ast$ 
induces the decomposition of the matrix $\Phi: A\bito B$ into
\begin{itemize}
\item the projection  $P^\ast:A \bito \cuu \Phi$ with $\mol {P^\ast} x {<\dow \alpha, \up \beta>} = x \dow \alpha$, and
\item the embedding $E^\ast:\cuu \Phi \bito B$ with $\mol {E^\ast} {<\dow \alpha, \up \beta>} y = \up \beta y$ 
\end{itemize}
where $<\dow \alpha, \up \beta>\in \cuu\Phi$ is an arbitrary $\Phi$-cut, i.e. $\dow \alpha = \Phi_\ast\up \beta$ and $\Phi^\ast \dow \alpha = \up \beta$. 
\end{thrm}

\bprf{ (sketch)} We prove that $\Phi = P^\ast ; E^\ast$ as follows:
\bear
\mol {(P^\ast; E^\ast)} x y & = &  \bigwedge_{\dow \alpha} \mol {P^\ast} x {\left<\dow\alpha, \Phi^\ast \dow \alpha\right>} +  \mol E {\left<\dow\alpha, \Phi^\ast \dow \alpha\right>} y \\
& = & \bigwedge_{\dow \alpha} x \dow \alpha +  \left(\Phi^\ast \dow \alpha\right) y \\
& \leq & x\mnd x + \left(\Phi^\ast \mnd x\right) y\\
& = & \dist A x x + \dist {\Up B} {\Phi^\ast \mnd x} {\cmn y} \\
& = &  \mol \Phi x y
\eear
using Lemma~\ref{repres-matrix-lemma} at the last step. The facts that $P^\ast$ is a projection and $E^\ast$ is an embedding matrix are proved using the following lemma, which says that $\cuu \Phi$ is $\tcoprod$-generated by $A$ and $\tprod$-generated by $B$.
\epr

\begin{lemm}
The $\cuu \Phi$-infima are computed in $\Do A$, whereas its suprema are computed in $\Up B$. To state this precisely, consider $\up \lambda : \cuu \Phi \to [0,\infty]$ and $\dow \varrho: \op{\cuu \Phi} \to [0,\infty]$. Extend them along the isomorphisms $\cuu\Phi_A \cong \cuu\Phi \cong \cuu \Phi_B$ to get 
$\up \lambda : \cuu\Phi_A \to [0,\infty]$ and $\dow \varrho: \op{\cuu\Phi_B} \to [0,\infty]$. Then
\[
\prod \up \lambda\  =\  \up \lambda \circ \mnd\   \in\  \cuu \Phi_A \qquad   \qquad \qquad \coprod \dow \varrho\  =\  \dow \varrho \circ \cmn\   \in\  \cuu \Phi_B
\]
are constructed in $\Do A$ and $\Up B$, because $\cuu\Phi_A \inclusion \Do A$ preserves the infima, whereas $\cuu\Phi_B \inclusion \Up B$ preserves the suprema.
\end{lemm}

\begin{corr}
The monotone maps $A\tto\mnd \Do A \eepi{p^\ast} \cuu \Phi \ipee{p_\ast} \Up B \oot \cmn B$ 
\begin{itemize}
\item preserve any infima that exist in $A$, and any suprema that exist in $B$, 
\item generate $\cuu \Phi$ by the suprema from $A$ and by the infima from $B$, in the sense that for any $<\dow \alpha, \up \beta>\in \cuu \Phi$ holds
\[ \coprod_{\dow \alpha} \mnd \ = \ <\dow \alpha, \up \beta>\ = \ \prod_{\up \beta} \cmn
\]
\end{itemize}
\end{corr}

\section{Bicompletion}\label{Semantic-compl-sec}
Any distance space morphism $f:A\to B$ induces two matrices, $\ohm f :A\bito B$ and $\mho f : B\bito A$ with
\[
\mol {\ohm f} x y = \dist B {fx} y \qquad \qquad \mol{\mho f} y x = \dist B y {fx}
\]

\begin{lemm}
For every matrix $\ohm f:A\bito B$ induced by a distance space morphism $f:A\to B$ holds $\dual {\ohm f} = \mho f$.
\end{lemm}

\bpr Since $\mol{\mho f} y x = \dist B y {fx}$ by definition, the claim boils down to $\mol {({\ohm f})^o} y x = \dist B y {fx}$, which can be proved as follows
\bear 
\mol {(\ohm f)^o} y x & = &  
\bigvee_{\substack{u\in A\\v\in B}} \dist B {fu} v \multimap 
\left( \dist B y v + \dist A u x\right) \\
& \geq & \dist B {fx} {fx} \multimap \left( \dist B y {fx} + \dist A x x\right)\ =\ \dist B y {fx} \eear
\epr

\subsection{Nucleus as a completion}
\begin{lemm}
If the distance space $B$ is complete, then for any matrix $\Phi:A\bito B$ there is a distance space morphism $f:A\to B$ such that $\Phi = \ohm f$.
\end{lemm}

\begin{corr}
If both $A$ and $B$ are complete, then any matrix $\Phi: A\bito B$ corresponds to an adjunction $\Phi^\ast :A\rightleftarrows B: \Phi_\ast$ such that $\Phi = \ohm \Phi^\ast = \mho \Phi_\ast$.
\end{corr}

\begin{defn}
A \emph{distance matrix homomorphism} $h:\Phi \to \Gamma$ where $\Phi: A\bito B$ and $\Gamma: C\bito D$, is a pair  of contractions $h = \left<h_0:A\to C, h_1:B\to D\right>$ such that
\begin{itemize}
\item $\ohm h_0\, ; \Gamma = \Phi\, ; \ohm h_1$,
\item $h_0$ preserves any suprema that may exist in $A$,
\item $h_1$ preserves any infima that may exist in $B$.
\end{itemize}
Let $\MMatr$ denote the category of distance space matrices and matrix morphisms. \end{defn}

\be{defn}
A matrix $\Phi: A\bito B$ is \emph{complete} if $A$ has suprema and $B$ infima\footnote{By Corollary~\ref{sup-inf}, both $A$ and $B$ are thus complete}, and $\Phi: \op{A} \times B \to [0,\infty]$ preserves the infima. Let $\CMatr$ denote the category of complete matrices and matrix homomorphisms.
\ee{defn}

\begin{proposition}\label{universal}
$\Id_{\cuu \Phi}: \cuu\Phi \bito \cuu \Phi$ is the completion of $\Phi:A\bito B$. In other words, the functor $\cuu{-} : \MMatr\to \CMatr$ is left adjoint to the full inclusion $\CMatr \hookrightarrow \MMatr$. The unit of the adjunction $\eta = <\eta_0, \eta_1>:\Phi \to  {\cuu \Phi}$ consists of 
\[ \eta_0: A\tto \mnd \Do A \tto {p^\ast} \cuu \Phi \quad \mbox{ and } \quad \eta_1: B\tto \cmn \Up B \tto {p_\ast} \cuu \Phi\]

\end{proposition}

\section{Summary and discussion}\label{Conclusions-sec}
Given an arbitrary distance matrix $\Phi: A\bito B$, we have constructed the completion $\Phi \tto \eta \cuu \Phi$ such that 
\begin{itemize}
\item $A\tto{\eta_0} \cuu \Phi$ is $\tcoprod$-generating and $\tprod$-preserving,
\item $B\tto{\eta_1} \cuu \Phi$ is $\tprod$-generating and $\tcoprod$-preserving.
\end{itemize}
In terms of the motivating example of program transformations, and of the task of conjoining the algorithmic knowledge about systems and about attacks, every $\Phi$-cut is thus a supremum of the system  specifications in $A$, and an infimum of the attack specifications in $B$. Moreover, the suprema of $\Phi$-cuts can be computed in  $\Up B$, whereas the infima can be computed in $\Do A$. While the suprema\footnote{not unlike colimits of software specifications \cite{SmithD:colimits,PavlovicD:SDR}} capture composite systems validating some composite properties, the infima describe composite attacks where the \emph{in}\/validated properties add up.

But what has been achieved by providing this very abstract account? It turns out that the actual completions provide fairly concrete information. There is no space to illustrate this, but we sketch a high level view. The prior knowledge,  represented by the distance spaces $A$ and $B$ is updated by the empiric data, represented by the matrix $\Phi: A\bito B$. In the completion $\cuu \Phi$, the empiric relations of $a$s and $b$s are expressed as distances. Following \cite[Ch.~4]{SolomonoffR:inference,Vitanyi:book}, the task of \emph{explaining} these empiric links can then be viewed as the task of finding short programs $p$ with $p(a) = b$. After such completions, some distances previously recorded in $A$ and $B$ may increase, since some programs may be closer related \emph{a posteriori}\/  than \emph{a priori}. 

The obvious task for future work is to refine the concrete applications of the presented construction. This is to some extent covered in the full paper, which is in preparation. The further work on quantifying the hardness of program derivations, and of program transformations, branches in many directions. Distances arise naturally in this framework, as described already in \cite[Sec.~4.2]{PavlovicD:NSPW11}. In a different direction, it seems interesting to study the bicompletions in other categorical frameworks, in particular where the dualities fail in a significant way, as demonstrated a long time ago \cite{IsbellJ:subobjects}.

\bibliographystyle{plain}
\bibliography{ref-semcom,PavlovicD}

\end{document}

%% file: macros.tex
\usepackage{pstricks}
\usepackage{amssymb}
\usepackage{amstext}
\usepackage{amsmath}
\usepackage{txfonts}
\usepackage{cmll}
\usepackage{txfonts}
\usepackage{rotating}
\usepackage{array}
\usepackage{pifont}

\usepackage{xy}
\xyoption{all}

\include{prooftree}

\spnewtheorem*{remarkno}{Remark}{\it}{\rmfamily}
\spnewtheorem*{explanationo}{Explanation}{\it}{\rmfamily}
\spnewtheorem*{propositiono}{Proposition}{\bf}{\it}
\spnewtheorem*{corollaryno}{Corollary}{\bf}{\it}
\spnewtheorem*{lemmano}{Lemma}{\bf}{\it}
\spnewtheorem{thrm}[proposition]{Theorem}{\bf}{\it}
\spnewtheorem{corr}[proposition]{Corollary}{\bf}{\it}
\spnewtheorem{lemm}[proposition]{Lemma}{\bf}{\it}
\spnewtheorem{defn}[proposition]{Definition}{\bf}{\it}

\newcounter{countroman}
{\begin{list}{{\rm (\roman{countroman})}}{\usecounter{countroman}}}%
{\end{list}}

\newcounter{countalpha}
\newenvironment{anumerate}%
{\begin{list}{(\alph{countalpha})}{\usecounter{countalpha}}}%
{\end{list}}

\newcounter{countalphabf}
{\protect\begin{list}{{\rm (}{\bf \protect\alph{countalphabf}}{\rm%
)}}{\protect\usecounter{countalphabf}}}%
{\end{list}}

 \newcommand{\cuu}[1]{\Lbag{#1}\Rbag}

\newcommand{\Dist}{{\sf Dist}}

\newcommand{\Matr}{{\sf Matr}}
\newcommand{\MMatr}{{\sf MMat}}
\newcommand{\CMatr}{{\sf CMat}}

\newcommand{\Id}{{\rm Id}}
\newcommand{\Leftt}{\Lambda}
\newcommand{\Rightt}{\Upsilon}
\newcommand{\Forgg}{{\sf W}}

\renewcommand{\to}{\rightarrow}

\newcommand{\tto}[1]{\xrightarrow{#1}}
\newcommand{\oot}[1]{\xleftarrow{#1}}

\newcommand{\epi}{\twoheadrightarrow}
\newcommand{\ipe}{\twoheadleftarrow}
\newcommand{\eepi}[1]{\stackrel{#1}{\twoheadrightarrow}}
\newcommand{\ipee}[1]{\stackrel{#1}{\twoheadleftarrow}}
\newcommand{\inclusion}{\hookrightarrow}
\newcommand{\clusionin}{\hookleftarrow}
\newcommand{\bito}{\looparrowright}

\newcommand{\dis}[2]{d\left({#1}  , {#2}\right)}
\newcommand{\dist}[3]{d_{#1}\left( {#2} ,  {#3}\right)}   

\newcommand{\mol}[3]{{#2}{#1}{#3}}

\newcommand{\WP}{\mbox{\Large $\wp$}}

\newcommand{\Pos}{{\sf Pos}}

\newcommand{\id}{{\rm id}}

\newcommand{\XXX}{{\cal X}}

\renewcommand{\Bbb}{\mathbb}

\newcommand{\NNn}{{\Bbb N}}

\mathcode`\<="4268 
\mathcode`\>="5269 
\mathchardef\gt="313E 
\mathchardef\lt="313C 

 \def\pushright#1{{
    \parfillskip=0pt            
    \widowpenalty=10000         
    \displaywidowpenalty=10000  
    \finalhyphendemerits=0      
    \leavevmode                 
    \unskip                     
    \nobreak                    
    \hfil                       
    \penalty50                  
    \hskip.2em                  
    \null                       
    \hfill                      
    {#1}                        
    \par}}                      

 \def\qed{\pushright{$\square$}\penalty-700 \smallskip}

\newenvironment{prf}[1]{\begin{trivlist} \item[{\bf ~Proof}#1.]}%
{\qed\end{trivlist}}

\newcommand{\be}[1]{\begin{#1}}

\newcommand{\ee}[1]{\end{#1}}
\newcommand{\beq}{\begin{equation}}
\newcommand{\eeq}{\end{equation}}
\newcommand{\ba}[1]{\begin{array}{#1}}
\newcommand{\ea}{\end{array}}
\newcommand{\bea}{\begin{eqnarray}}
\newcommand{\eea}{\end{eqnarray}}
\newcommand{\bear}{\begin{eqnarray*}}
\newcommand{\eear}{\end{eqnarray*}}
\newcommand{\bpr}{\begin{prf}{}}
\newcommand{\epr}{\end{prf}}
\newcommand{\bprf}[1]{\begin{prf}{#1}}
\newcommand{\eprf}{\end{prf}}

\renewcommand{\paragraph}[1]{\smallskip\noindent{\bf #1} }

\newcommand{\op}[1]{{#1}^{o}}
\newcommand{\dual}[1]{{#1}^\ddag}

\newcommand{\Up}[1]{\overleftarrow{#1}}
\newcommand{\Do}[1]{\overrightarrow{#1}}
\newcommand{\UD}[1]{\overleftrightarrow{#1}}

\newcommand{\up}[1]{\overleftarrow{#1}}

\newcommand{\dow}[1]{\overrightarrow{#1}}

\newcommand{\cmn}{{\rm \Delta}}
\newcommand{\mnd}{\nabla}

\newcommand{\tprod}{\textstyle\prod}
\newcommand{\tcoprod}{\textstyle\coprod}
\newcommand{\ohm}{{\rm \Omega}}

\newcommand{\Dashv}{\mathrel{\mbox{$={\kern-1.1ex}|$}}}
\renewcommand{\vDash}{\mathrel{\mbox{$|{\kern-1.1ex}=$}}}

\newcommand{\closeto}{\leadsto}

%% file: prooftree.tex
\newdimen\proofrulebreadth \proofrulebreadth=.05em
\newdimen\proofdotseparation \proofdotseparation=1.25ex
\newdimen\proofrulebaseline \proofrulebaseline=2ex
\newcount\proofdotnumber \proofdotnumber=3
\let\then\relax
\def\hfi{\hskip0pt plus.0001fil}
\mathchardef\squigto="3A3B
\newif\ifinsideprooftree\insideprooftreefalse
\newif\ifonleftofproofrule\onleftofproofrulefalse
\newif\ifproofdots\proofdotsfalse
\newif\ifdoubleproof\doubleprooffalse
\let\wereinproofbit\relax
\newdimen\shortenproofleft
\newdimen\shortenproofright
\newdimen\proofbelowshift
\newbox\proofabove
\newbox\proofbelow
\newbox\proofrulename
\def\shiftproofbelow{\let\next\relax\afterassignment\setshiftproofbelow\dimen0 }
\def\shiftproofbelowneg{\def\next{\multiply\dimen0 by-1 }%
\afterassignment\setshiftproofbelow\dimen0 }
\def\setshiftproofbelow{\next\proofbelowshift=\dimen0 }
\def\setproofrulebreadth{\proofrulebreadth}

\def\prooftree{
\ifnum  \lastpenalty=1
\then   \unpenalty
\else   \onleftofproofrulefalse
\fi
\ifonleftofproofrule
\else   \ifinsideprooftree
        \then   \hskip.5em plus1fil
        \fi
\fi
\bgroup
\setbox\proofbelow=\hbox{}\setbox\proofrulename=\hbox{}%
\let\justifies\proofover\let\leadsto\proofoverdots\let\Justifies\proofoverdbl
\let\using\proofusing\let\[\prooftree
\ifinsideprooftree\let\]\endprooftree\fi
\proofdotsfalse\doubleprooffalse
\let\thickness\setproofrulebreadth
\let\shiftright\shiftproofbelow \let\shift\shiftproofbelow
\let\shiftleft\shiftproofbelowneg
\let\ifwasinsideprooftree\ifinsideprooftree
\insideprooftreetrue
\setbox\proofabove=\hbox\bgroup$\displaystyle 
\let\wereinproofbit\prooftree
\shortenproofleft=0pt \shortenproofright=0pt \proofbelowshift=0pt
\onleftofproofruletrue\penalty1
}

\def\eproofbit{
\ifx    \wereinproofbit\prooftree
\then   \ifcase \lastpenalty
        \then   \shortenproofright=0pt  
        \or     \unpenalty\hfil         
        \or     \unpenalty\unskip       
        \else   \shortenproofright=0pt  
        \fi
\fi
\global\dimen0=\shortenproofleft
\global\dimen1=\shortenproofright
\global\dimen2=\proofrulebreadth
\global\dimen3=\proofbelowshift
\global\dimen4=\proofdotseparation
\global\count255=\proofdotnumber
$\egroup  
\shortenproofleft=\dimen0
\shortenproofright=\dimen1
\proofrulebreadth=\dimen2
\proofbelowshift=\dimen3
\proofdotseparation=\dimen4
\proofdotnumber=\count255
}

\def\proofover{
\eproofbit 
\setbox\proofbelow=\hbox\bgroup 
\let\wereinproofbit\proofover
$\displaystyle
}%
\def\proofoverdbl{
\eproofbit 
\doubleprooftrue
\setbox\proofbelow=\hbox\bgroup 
\let\wereinproofbit\proofoverdbl
$\displaystyle
}%
\def\proofoverdots{
\eproofbit 
\proofdotstrue
\setbox\proofbelow=\hbox\bgroup 
\let\wereinproofbit\proofoverdots
$\displaystyle
}%
\def\proofusing{
\eproofbit 
\setbox\proofrulename=\hbox\bgroup 
\let\wereinproofbit\proofusing
\kern0.3em$
}

\def\endprooftree{
\eproofbit 
  \dimen5 =0pt
\dimen0=\wd\proofabove \advance\dimen0-\shortenproofleft
\advance\dimen0-\shortenproofright
\dimen1=.5\dimen0 \advance\dimen1-.5\wd\proofbelow
\dimen4=\dimen1
\advance\dimen1\proofbelowshift \advance\dimen4-\proofbelowshift
\ifdim  \dimen1<0pt
\then   \advance\shortenproofleft\dimen1
        \advance\dimen0-\dimen1
        \dimen1=0pt
        \ifdim  \shortenproofleft<0pt
        \then   \setbox\proofabove=\hbox{%
                        \kern-\shortenproofleft\unhbox\proofabove}%
                \shortenproofleft=0pt
        \fi
\fi
\ifdim  \dimen4<0pt
\then   \advance\shortenproofright\dimen4
        \advance\dimen0-\dimen4
        \dimen4=0pt
\fi
\ifdim  \shortenproofright<\wd\proofrulename
\then   \shortenproofright=\wd\proofrulename
\fi
\dimen2=\shortenproofleft \advance\dimen2 by\dimen1
\dimen3=\shortenproofright\advance\dimen3 by\dimen4
\ifproofdots
\then
        \dimen6=\shortenproofleft \advance\dimen6 .5\dimen0
        \setbox1=\vbox to\proofdotseparation{\vss\hbox{$\cdot$}\vss}%
        \setbox0=\hbox{%
                \advance\dimen6-.5\wd1
                \kern\dimen6
                $\vcenter to\proofdotnumber\proofdotseparation
                        {\leaders\box1\vfill}$%
                \unhbox\proofrulename}%
\else   \dimen6=\fontdimen22\the\textfont2 
        \dimen7=\dimen6
        \advance\dimen6by.5\proofrulebreadth
        \advance\dimen7by-.5\proofrulebreadth
        \setbox0=\hbox{%
                \kern\shortenproofleft
                \ifdoubleproof
                \then   \hbox to\dimen0{%
                        $\mathsurround0pt\mathord=\mkern-6mu%
                        \cleaders\hbox{$\mkern-2mu=\mkern-2mu$}\hfill
                        \mkern-6mu\mathord=$}%
                \else   \vrule height\dimen6 depth-\dimen7 width\dimen0
                \fi
                \unhbox\proofrulename}%
        \ht0=\dimen6 \dp0=-\dimen7
\fi
\let\doll\relax
\ifwasinsideprooftree
\then   \let\VBOX\vbox
\else   \ifmmode\else$\let\doll=$\fi
        \let\VBOX\vcenter
\fi
\VBOX   {\baselineskip\proofrulebaseline \lineskip.2ex
        \expandafter\lineskiplimit\ifproofdots0ex\else-0.6ex\fi
        \hbox   spread\dimen5   {\hfi\unhbox\proofabove\hfi}%
        \hbox{\box0}%
        \hbox   {\kern\dimen2 \box\proofbelow}}\doll%
\global\dimen2=\dimen2
\global\dimen3=\dimen3
\egroup 
\ifonleftofproofrule
\then   \shortenproofleft=\dimen2
\fi
\shortenproofright=\dimen3
\onleftofproofrulefalse
\ifinsideprooftree
\then   \hskip.5em plus 1fil \penalty2
\fi
}
